\documentstyle[psfig]{mn}
\def\H0{{\it H}$_0$}
\def\Ms{{\it M}$_\odot$}

\def\q0{{\it q}$_0$}
\def\kmps{km~s$^{-1}$}

\def\ergps{erg~s$^{-1}$}
\def\kmpspMpc{km~s$^{-1}$~Mpc$^{-1}$}
\def\Ms{{\it M}$_\odot$}

\def\nH{$N_{\rm H}$\thinspace} 
\def\DelnH{$\Delta N_{\rm H}$}
\def\psqcm{cm$^{-2}$}
\def\ergpspsqcm{erg~cm$^{-2}$~s$^{-1}$}

\def\Zs{$Z_{\odot}$}

\def\cps{ct\thinspace s$^{-1}$}

\def\phpspsqcm{ph\thinspace s$^{-1}$\thinspace cm$^{-2}$}

\def\cubcm{cm$^{-3}$}

\title[X-ray emission from 4C+55.16] 
{X-ray and lensing results on the cluster around the powerful radio galaxy 4C+55.16} 
\author[K. Iwasawa et al] 
{\parbox[]{6.5in} {K. Iwasawa$^1$, S.W. Allen$^1$, A.C. Fabian$^1$, A.C. Edge$^2$ and S. Ettori$^1$}\\
\\
$^1$ Institute of Astronomy, Madingley Road, Cambridge CB3 0HA\\ 
$^2$ Department of Physics, University of Durham, South Road, Durham DH1 3LE\\}

\date{}

\begin{document}

\maketitle

\begin{abstract}
We present results from ASCA and ROSAT HRI observations of 
the powerful radio galaxy 4C+55.16
at redshift 0.24. Extended soft X-ray emission is imaged
by the ROSAT HRI. The X-ray brightness profile is sharply peaked on the 
radio galaxy, characteristic of a strong cooling flow.
The X-ray spectrum obtained from ASCA 
is consistent with multi-phase intracluster gas.
There is evidence for an absorbed cool component as well as ambient
cluster medium in the ASCA spectrum.
A spectral fit, taking a cooling-flow component into account, gives 
a temperature of $kT = 5.4^{+1.4}_{-0.9}$ keV, metal abundance 
$(0.5\pm 0.1)$\Zs, excess absorption on the cool component, 
$\Delta N_{\rm H} = 4.9^{+3.4}_{-1.3}\times 10^{21}$\psqcm, and 
an absorption-corrected bolometric luminosity of $2.2\times 10^{45}$\ergps\
(\H0\ = 50 \kmpspMpc, \q0\ = 0.5).
The mass deposition rate is estimated to be $\sim 1100$\Ms\ yr$^{-1}$ from the 
spectral analysis, in good agreement with that derived from 
imaging analysis of the ROSAT data when corrected for absorption.
We tentatively identify a blue feature, seen $\sim 15$ arcsec
from the centre of the radio galaxy in a published optical image, 
as a gravitationally-lensed arc. 
The inferred lensing mass is consistent with the gravitational
mass derived from the X-ray data. The best-estimate of the redshift of
the lensed object is $\sim 1.5 (>0.7)$.
All the observed properties suggest that the environment of 4C+55.16
is similar to known massive cooling flow clusters.
This is the first massive cooling flow to be found around a
powerful, radio source with a compact, GHz-peaked spectrum core.
\end{abstract}

\begin{keywords}
galaxies: clusters: individual: 4C+55.16 --
X-rays: galaxies --
cooling flows --
gravitational lensing
\end{keywords}

\section{Introduction}

4C+55.16 (= 0831+557) is a radio galaxy at a redshift of 0.240 
(Lawrence et al 1996) in the 
Pearson-Readhead (1981, 1988) 5 GHz flux-density-limited complete sample.
The radio source is compact and powerful 
($1.1\times 10^{26}$W\thinspace Hz$^{-1}$sr$^{-1}$ at 5 GHz).
In the high resolution radio maps obtained from MERLIN 
and VLBI (Whyborn et al 1985; Pearson \& Readhead 1988),
highly irregular, small-scale structures have been resolved.
The radio spectrum of the compact core shows a turnover 
around 1 GHz (Component C in Whyborn et al 1985),
resembling the class of Giga-hertz Peaked Spectrum (GPS) sources
(e.g., O'Dea 1998) whilst the mini double-lobe with a projected size 
of 11\arcsec\ (54 kpc at a redshift of 0.24) has 
a steep, low-frequency spectrum.
A sum of the two components results in a flat radio spectrum 
over the GHz range.

A blue continuum contributes $\sim 50$ per cent 
of the light at 3850\AA\ (Heckman et al 1983) 
and its optical emission-line spectrum 
%taken by the Palomer 200 inch Spectrograph 
shows a medium ionization state 
(e.g., Heckman et al 1983 based on the [OII]$\lambda 5007$/[OII]$\lambda 3727$
ratio; Whyborn et al 1985;
[OIII]$\lambda 5007$/H$\beta\sim 4$, Lawrence et al 1996).

X-ray emission was detected during the ROSAT All Sky Survey
(RX\thinspace J08349+5534, Brinkmann et al 1995; Laurent-Muehlei et al 1996;
Bade et al 1998).
The ROSAT papers assumed that 
the X-ray emission originates in an active nucleus residing in 4C+55.16.
However, as our imaging and spectral study using the ROSAT HRI and ASCA data
show below, the observed X-rays appear to 
be dominated by cluster emission surrounding the radio galaxy.

The optical CCD images of 4C+55.16
were obtained by Hutchings, Johnson \& Pyke (1988)
using the Canada-France-Hawaii Telescope (CFHT) with $B$ and $R$ filters.
% The radio galaxy appears to be a large, central galaxy in a compact 
% galaxy cluster.
There is a blue feature about 15 arcsec southwest 
from the centre of the galaxy, which we tentatively identify
as a gravitationally lensed arc.
We find a good agreement between the cluster masses estimated from
X-ray and lensing techniques.

%The optical position of the radio galaxy is 8h34m54.9s, +55d34m21s19.0.

\section{Observations and data reduction}

% Observation Summary  --- Table 1

\begin{table*}
\begin{center}
\caption{The ASCA and ROSAT observations of 4C+55.16. No corrections for vignetting has been made for the count rates given here.}
\begin{tabular}{lcccccc}
Satellite & Detector & Operation mode & Band & Date & Exposure & Count rate \\
          &          &                &      &      & ks & \cps \\[5pt]
ASCA &    SIS (S0/S1) & 1CCD/Faint & 0.5--10 keV &  1996 Nov 1--2  &  
34.5/34.0 & 0.12/0.10 \\
     &    GIS (G2/G3) & PH    &     0.7--10 keV & & 38.1/37.8 & 0.065/0.078 \\[5pt]
ROSAT &   HRI & &         0.1--2.4 keV & 1997 Oct 20 & 16.5 & 0.062 \\
\end{tabular}
\end{center}
\end{table*}

4C+55.16 was observed with ASCA (Tanaka, Inoue \& Holt 1994) and ROSAT HRI
(Pfeffermann et al 1987).
ASCA provides moderate resolution imaging and spectra in the 
0.5--10 keV band while the ROSAT HRI provides high resolution 
(FWHM $\sim$ 5 arcsec) 0.1--2.4 keV imaging.
A summary of the observations is given in Table 1.

The four detectors onboard ASCA, the Solid state Imaging Spectrometers 
(SIS; S0 and S1), the Gas Imaging Spectrometers (GIS; G2 and G3) were
operating normally. 
The radio galaxy was pointed at the nominal position on the
best-calibrated CCD chip of each SIS detector.
The lowest energy event threshould in the SIS was set at 0.47 keV.
Standard calibration (used for the Rev2 processing) and 
data reduction techniques were employed,
using FTOOLS (version 4.0) provided by the ASCA Guest Observer Facility
at Goddard Space Flight Center.
The SIS data were corrected for the effects of the Dark Frame Error (DFE)
and `Echo' and hot/flickering pixels were removed.

The ROSAT HRI observation was short (16.6 ks).
The soft X-ray image revealed extended X-ray emission which could not be 
resolved with ASCA (Section 3). 
The data were also analysed by the deprojection technique
(Section 5). There is no evidence for X-ray variation in both ASCA and ROSAT
data.

\section{The ROSAT HRI image}

% HRI image on DSS --- Fig. 1

\begin{figure}
%\vspace{6cm}
%\centerline{\psfig{figure=/data/ki/4c55/hri/idl_overlay.ps,width=0.42\textwidth,angle=0}}
\centerline{\psfig{figure=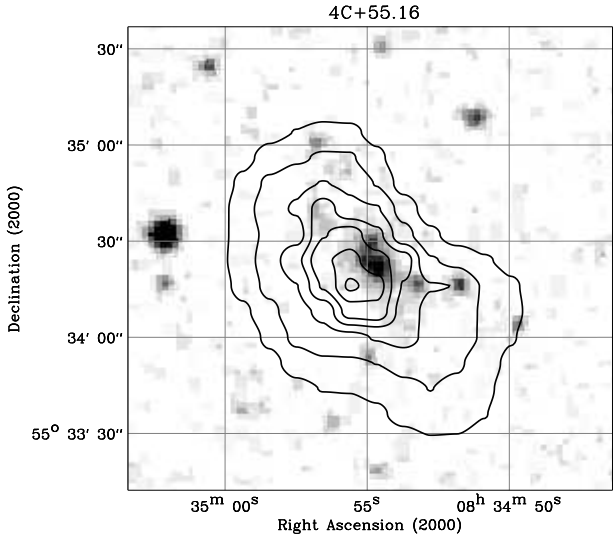,width=0.45\textwidth,angle=0}}
\caption{The X-ray (0.1--2.4 keV) contours of 4C+55.16 obtained from the 
ROSAT HRI image overlaied onto the optical DSS image taken by Palomer 48-inch 
Oschin Schmidt Telescope. The DSS image is not deep enough to show the 
arc-like feature detected in the CFHT image (Hutchings et al 1988). 
Contour levels are 6, 10, 16, 25, 40, 63, 95 per cent of the peak brightness
which is 63 ct pixel$^{-1}$. The estimated average background is about 
1.3 ct pixel$^{-1}$.}
\end{figure}

The X-ray source detected at the position of 4C+55.16 is clearly extended
beyond the Point Spread Function (PSF) of the HRI.
The raw HRI image has been smoothed with the adaptive kernel
method, ASMOOTH (Ebeling, White \& Rangarajan 1998),
using a gaussian kernel and a characteristic smoothing threshold,
above the local background, of $4\sigma$.
Fig. 1 sows the
X-ray intensity contours of the smoothed data overlaid on the
optical Digitized Sky Survey (DSS) image. 

The X-ray emission is sharply peaked at the centre of the extended image,
as observed in strong cooling flow clusters.
The X-ray peak is displaced from the galaxy nucleus by $\sim 7$ arcsec,
well within the mean displacement (25 arcsec) 
between X-ray and optical peaks seen
in cooling flow clusters observed with ROSAT HRI (Peres et al 1997).
Although the displacement is consistent with the positional uncertainty
of ROSAT pointing ($\sim 10$ arcsec), if it is real the galaxy resides 
in the X-ray cavity, as seen in the Perseus cluster (B\"ohringer et al 1993).
There are three point-like sources detected in the HRI field.
%However, since they have no obvious optical counterpart in the DSS image,
%it is not possible to check the astrometry.
One of them shows an offset from a possible optical counterpart,
similar in amplitude and direction to that for 4C+55.16. 
Therefore the displacement between
X-ray and optical peaks may be merely due to a pointing error.

The elongation of the X-ray image is prominent at the low suface 
brightness levels. However, this may be partly due to a contamination
from point sources in the field, particularly in the SW region.
The present observation is too short to investigate such a faint
X-ray morphology.

\section{The ASCA X-ray spectrum}

The spectral analysis was performed using XSPEC (version 10.0).
The MEKAL model (the original MEKA code, described by Kaastra 1992, 
with modified Fe-L line emissivity by Liedahl et al 1995)
for an optically thin, collisional ionization equilibrium 
plasma was used with solar abundances taken from Anders \& Grevesse (1989).
The photoelectric absorption model was taken from Morrison \& McCammon (1983).
The Galactic absorption at the position of 4C+55.16
is estimated to be \nH\ = $4.2\times 10^{20}$\psqcm\ from the
HI measurements of Dickey \& Lockman (1990). Absorption column densities
obtained from the spectral fits are then excesses above the Galactic
value.
Quoted errors to the best-fit spectral parameters are 90 per cent confidence
regions for one parameter of interest.

%Thermal X-ray emission

The ASCA spectrum shows a clear line feature at 5.4 keV, which
is in an excellent agreement with the rest Fe K$\alpha $ line emission at
$\sim 6.7$ keV expected from a thin thermal plasma with a
temperature of several keV at the redshift ($z = 0.240$) of 4C+55.16.
The agreement between the redshifts of the Fe K$\alpha $ 
line and the radio galaxy
strongly supports the presence of a cluster around the radio galaxy
rather than in the background which was suspected by
Hutchings et al (1988).
The thermal emission model (MEKAL) provides a slightly better fit
to the data (0.6--9 keV from the SIS; 0.9--10 keV from the GIS)
than the model of a power-law plus a gaussian line for the Fe K$\alpha$
line feature (see Table 2).
The strong, narrow Fe K$\alpha$ line at 6.7 keV 
is unlikely for an active galaxy
but naturally explained by thermal emission from cluster gas whose 
X-ray emission has been spatially resolved by the ROSAT HRI.

\begin{table*}
\begin{center}
\caption{Spectral fits to the ASCA SIS (0.6--9 keV) and GIS (0.9--10 keV)
data. \DelnH\ is the excess column density above the
Galactic value ($4.2\times 10^{20}$\psqcm), measured 
in the galaxy rest frame. Errors are quoted at the 90 per cent 
confidence level for one parameter of interest. In the model (a), 
the gaussian line energy is corrected for the galaxy redshift. 
The equivalent width of the line is $520^{+144}_{-138}$ eV. 
In the thermal emission model ({\tt MEKAL}), the Solar abundance table
from Anders \& Grevesse (1989) is used. In the model (c), 
the cool component is the cooling flow model with
an initial temperature $kT_i$, metal abundane $Z$ and a mass deposition 
rate of \.M. The column density of cold absorption 
for the cool component ($\Delta N_{\rm H, cool}$) is measured at $z=0.24$.}
\begin{tabular}{lccccc}
\multicolumn{6}{c}{(a) Power-law plus a Gaussian line} \\[10pt]
$\Gamma$ & \DelnH & $E$ & $\sigma$ & $I$ & $\chi^2$/dof \\
& $10^{21}$\psqcm & keV & keV & $10^{-5}$\phpspsqcm & \\[5pt]
$2.23^{+0.07}_{-0.05}$ & $3.24^{+0.54}_{-0.44}$ & $6.70^{+0.05}_{-0.06}$ &
$0 (<0.12)$ & $1.65^{+0.45}_{-0.43}$ & 529.6/511 \\[10pt]
\multicolumn{6}{c}{(b) Single thermal emission model (MEKAL)}\\[10pt]
$kT$ & $Z$ & \DelnH & & & $\chi^2$/dof \\
keV & \Zs & $10^{21}$\psqcm & & & \\[5pt]
 $4.06^{+0.27}_{-0.26}$ & $0.48^{+0.11}_{-0.10}$ & 
$1.28^{+0.38}_{-0.36}$ & & & 521.8/513 \\[10pt] 
\multicolumn{6}{c}{(c) Multi-phase thermal emission model}\\[10pt]
$kT_i$ & $Z$ & \.M & $\Delta N_{\rm H, cool}$ & & $\chi^2$/dof \\
keV & \Zs & \Ms yr$^{-1}$ & $10^{21}$\psqcm & & \\[5pt]
$5.4^{+1.4}_{-0.9}$ & $0.49^{+0.13}_{-0.12}$ & $1100^{+240}_{-410}$ &
$4.9^{+3.4}_{-1.3}$ & & 524.0/512 \\ 
\end{tabular}
\end{center}
\end{table*}

%Table A2

\begin{table*}
\begin{center}
\caption{Thermal emission model fits to the data above and below 3 keV.}
\begin{tabular}{ccccc}
Data & $kT$ & $Z$ & \DelnH & $\chi^2$/dof \\
& keV & \Zs & $10^{21}$\psqcm & \\[5pt]
E$>$3 keV & $5.40^{+1.01}_{-0.81}$ & $0.48^{+0.12}_{-0.11}$ & 
0  & 151.4/156\\ 
E$<$3 keV & $3.73^{+0.71}_{-0.46}$ & $0.59^{+0.37}_{-0.25}$ & 
$0.98^{+0.34}_{-0.32}$ & 406.7/394 \\ 
\end{tabular}
\end{center}
\end{table*}

%Multi-phase ICM

There is evidence for multi-phase gas in the ASCA spectrum.
Fits to the ASCA data above and below 3 keV with a single 
thermal emission model (MEKAL) give significantly different values of 
temperature (Table 3).
The data below 3 keV also require excess absorption 
($\Delta N_{\rm H}=0.98^{+0.34}_{-0.32}\times 10^{21}$\psqcm\
measured in the Earth frame) above the Galactic value.
The extrapolation of the best-fit model for the 3--10 keV data 
leaves excess emission down to 1 keV followed by a decline towards 0.6 keV
is probably due to absorption (Fig. 2). 
This indicates a multi-phase gas consisting of at least 
an absorbed, cool component 
with a less absorbed, ambient medium.
This is characteristic of a cooling flow.

% soft ratio: data vs extrapolation of hard band model

\begin{figure}
\centerline{\psfig{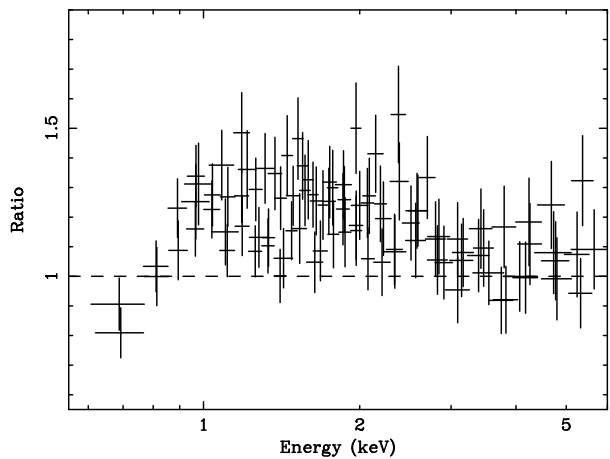}}
\caption{The ratio plot of the data and the extrapolation of the
MEKAL model best-fitting the 3--10 keV data. Excess soft X-ray emission
can be seen down to 1 keV and then it declines towards lower energies.}
\end{figure}

%Multi-phase model fit

A fit to the whole band data (0.6--8 keV for the SIS; 0.9--10 keV for
the GIS) with a multi-phase model (Model-c in Table 2), 
consisting of a single temperature MEKAL
with Galactic absorption and the cooling flow model (Johnstone et al 1992)
modified
by extra absorption at the source, gives temperature, 
metal abundance, mass deposition rate and excess absorption column for
the cooled component of $kT = 5.4^{+1.4}_{-0.9}$ keV, 
$Z=0.50^{+0.12}_{-0.13}$\Zs, \.M $= 1100^{+240}_{-410}$\Ms yr$^{-1}$,
and \nH $=4.9^{+3.4}_{-1.3}\times 10^{21}$\psqcm, respectively.
The temperature is listed for the ambient medium (i.e., before cooling),
the metallicity is assumed to be identical between the two components,
and the absorption column density is corrected for the galaxy redshift.
The quality of the fit is acceptable and comparable to 
the single phase model (see Table 2).
The covering fraction of the cold absorption must be larger than
0.9 (90 per cent lower limit).

The observed fluxes in the 0.5--2 keV and 2--10 keV bands 
obtained from the GIS are 
%$1.57\times 10^{-12}$\ergpspsqcm\ and     -- SIS value
%$2.33\times 10^{-12}$\ergpspsqcm, respectively. -- SIS value
$1.72\times 10^{-12}$\ergpspsqcm\ and
$2.56\times 10^{-12}$\ergpspsqcm, respectively.
In the best-fit multi-phase model, about half of the total flux 
comes from the cool component.

%Fe K beta anomaly

% SIS Fe K band spectrum

\begin{figure}
\centerline{\psfig{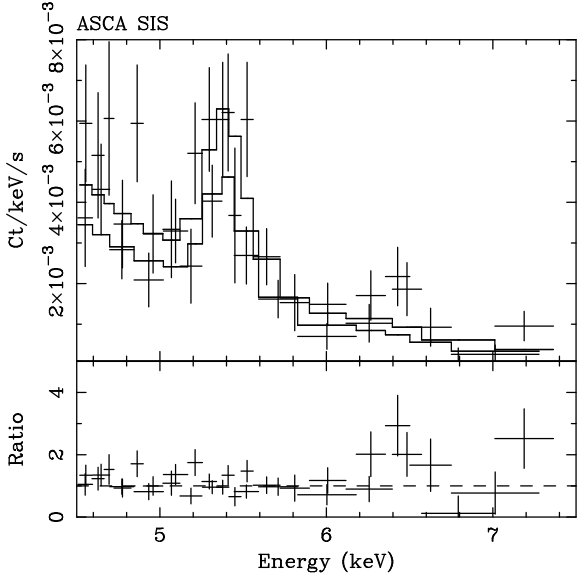}}
\caption{The Fe K band spectrum obtained from the ASCA SIS. The solid-line
histrogram represents the best-fit thermal emission model (See Table 3). 
Fe K$\beta$ emission around 6.4 keV (7.9 keV in the galaxy rest-frame)
is detected with  marginal significance above the model.}
\end{figure}

Excess line-like emission at a rest energy of 7.9 keV (6.4 keV
in the observed frame) is marginally detected in the SIS data (Fig. 3).
It can be identified with Fe K$\beta$. 
This feature is barely seen in the G3 data but not in the G2 data 
which show an unusual deficit between 6--7 keV, probably due to some anomaly
in the detector.
Since there is an instrumental feature at 6.4 keV in the SIS spectrum,
the SIS results should be treated with caution.
If the feature in the SIS data is real,
the Fe K$\beta$ emission is underestimated by a factor of $2.3\pm 1.5$ by the 
best-fit thermal emission model. 
An anomalous Fe K$\beta$/Fe K$\alpha$
ratio has been observed in the central region of a few cooling flow 
clusters, and has been 
interpreted as the effect of resonant scattering of line photons
in the cluster core (e.g., Akimoto et al 1996; Molendi et al 1998).
This interpretation is however not applicable to 4C+55.16 because the whole
cluster emission is observed. Further deeper observations are required 
to confirm the existence of this emission feature.

%\begin{figure*}
%\vspace{4cm}
%\caption{radial profile of X-ray emission,...}
%\end{figure*}

\section{Deprojection analysis of the HRI data and the X-ray mass model}

We have carried out a deprojection analysis of the ROSAT HRI data. 
An azimuthally-averaged 
X-ray surface brightness profile was constructed for the cluster. This was 
background-subtracted, corrected for telescope vignetting and re-binned into 
12 arcsec bins to provide sufficient counts in each annulus for a reliable 
statistical analysis to be carried out. With the X-ray surface brightness 
profile as the primary input, and under assumptions of spherical symmetry 
and hydrostatic equilibrium, the deprojection technique yields
the basic properties of the intracluster gas (temperature, density, pressure, 
cooling rate) as a function of radius.

The deprojection method requires the total mass profile for the 
cluster to be specified. We have iteratively determined the mass profile 
(which has been parameterized as an isothermal sphere; Equation 4-125 of 
Binney \& Tremaine 1987) that results in a deprojected temperature profile 
which is isothermal within the region probed by the ROSAT data and which 
is consistent with the best-fit temperatures determined from the ASCA 
spectra, using the cooling-flow model (Section 4, The validity of the 
assumption of isothermal mass-weighted temperature profiles in the cluster 
cores is discussed by Allen 1998b). The best-fitting mass model has 
a core radius of $60\pm20$ kpc and a velocity dispersion of 
$820^{+100}_{-70}$ \kmps.

The primary results from the deprojection analysis are as follows: for an
assumed Galactic column density of $4.2\times10^{20}$\psqcm, we determine 
the mean cooling time within the central 
12 arcsec bin of $t_{\rm cool}=2.0^{+0.3}_{-0.2} \times 10^9$ yr, a cooling 
radius (beyond which the cooling time exceeds a Hubble time) of 
$r_{\rm cool}=180^{+130}_{-40}$ kpc, and an integrated mass deposition rate 
within the cooling radius of ${\dot M}=460^{+260}_{-140}$\Ms\ yr$^{-1}$. If we 
correct for intrinsic absorption in the cluster, as determined from the ASCA 
data, these values are adjusted to $t_{\rm cool}=1.5^{+0.2}_{-0.2} \times 
10^9$ yr, $r_{\rm cool}=270^{+50}_{-10}$ kpc, and ${\dot
M}=970^{+270}_{-450}$\Ms\ yr$^{-1}$.

\section{Lensing analysis and comparison with the X-ray results}

The mass model determined from the X-ray data may be compared to
the mass implied by the observed lensing configuration in the cluster
(Section 1). Since only a single, putative gravitational arc is seen,
and to be consistent with the X-ray 
analysis, we have only carried out a simple, spherically-symmetric analysis 
of the lensing data. For a spherical mass distribution, the projected mass 
within the tangential critical radius, which we assume to be equal to the 
arc radius, $r_{\rm arc}= 15$ arcsec (71.8 kpc), is given by

\begin{equation}
M_{\rm arc}(r_{\rm arc}) ~ =
\frac{c^2 }{4 G} \left( \frac{D_{\rm arc}} {D_{\rm clus} D_{{\rm
arc-clus}}}
\right) ~ r_{\rm arc}^2
\end{equation}

\noindent where $D_{\rm clus}$, $D_{\rm arc}$ and $D_{\rm arc-clus}$ are
respectively the angular diameter distances from the observer to the 
cluster, the observer to the lensed object, and the cluster to the lensed 
object.  Fig. 4 shows the projected mass within the 
critical radius as a function of the redshift of the arc (solid curve). 
The horizontal dashed and dotted lines mark the best fit (projected) 
mass measurement and 90 per cent confidence limits determined
from the X-ray data, within the same radius ($3.8^{+1.0}_{-0.6} 
\times 10^{13}$ \Ms). We see that the X-ray and lensing mass measurements 
are consistent for any arc redshift $z_{\rm arc} > 0.7$. The best match 
between the X-ray and lensing mass measurements is obtained for an arc 
redshift of 1.5.

% Lensing mass --- Fig. 4

\begin{figure}
\centerline{\psfig{figure=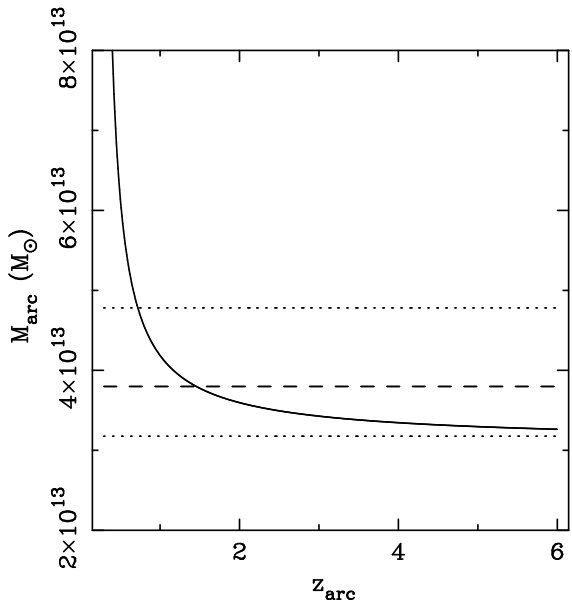,width=0.65\textwidth,angle=270}}
\caption{The projected lensing mass within the critical radius 
as a function of the redshift of the lensed object (solid curve). 
The gravitational mass estimated
from the deprojection anlaysis using the ASCA results 
is indicated by a dashed line with the 90 per cent
confidence limits shown by dotted lines.}
\end{figure}

\section{Discussion}

The ROSAT HRI image of the powerful radio galaxy 4C+55.16 shows 
extended X-ray emission peaking at the radio galaxy indicating 
cluster emission with a strong cooling flow.

A spectral study of the ASCA data suggests the X-ray emitting gas to be 
multi-phase. 
An absorbed, cool component is found in the spectrum.
The muti-phase spectral analysis indicates that the temperature of the 
ambient cluster medium is $kT \simeq 5.4$ keV.
A single-phase model fitted to the data gives a temperature lower by
$\sim 1$ keV, typical of a cooling flow cluster.

The mass deposition rate of the cooling flow,
$1100^{+240}_{-410}$\Ms\ yr$^{-1}$, derived from the spectral analysis is
consistent with that ($970^{+270}_{-450}$\Ms\ yr$^{-1}$) 
estimated from the image analysis when corrected for excess absorption.
Agreements between mass deposition rates derived from the two methods 
have been found for other distant cooling-flow clusters (Allen 1998b).

The optical spectrum of 4C+55.16 (Lawrence et al 1996) is indeed 
very similar to the other cooling flow galaxies (Crawford et al 1999).
The H$\alpha $ luminosity is about $8\times 10^{42}$\ergps\
(Lawrence et al 1996).
The relatively large Balmer decriment (H$\alpha $/H$\beta \simeq 5.5$)
suggests a significant reddening, which is 
often observed in cooling flows. The inferred absorption column density is 
slightly smaller than observed in the X-ray spectrum.

The absorption-corrected 2--10 keV and bolometric lumiosities, 
computed from the multi-phase model, are $8.0\times 10^{44}$\ergps,
and $2.2\times 10^{45}$\ergps, respectively (\H0\ = 50 \kmpspMpc\ and 
\q0\ = 0.5).
About 60 per cent of the bolometric luminosity is due to the cooling flow.
The bolometric luminosity exceeds that predicted for the single-phase 
temperature of 4 keV from the correlation between emission-weighted
cluster temperature and luminosity (Mushotzky 1984;
Edge \& Stewart 1991; David et al 1993;
Fabian et al 1994; Mushotzky \& Sharf 1997; White et al 1997).
A similar descrepancy is found for 
the other strong cooling flow clusters (e.g., Fabian et al 1994; 
Allen \& Fabian 1998a; Markevich 1998).
Taking the temperature derived from the multi-phase spectral analysis,
4C+55.16 fits well the $kT_{\rm X}$--$L_{\rm Bol}$ correlation obtained 
from a similar analysis of
other luminous ($L_{\rm Bol}>10^{45}$\ergps)
clusters for which the effect of cooling flows is included 
(Allen \& Fabian 1998a), and is consistent with 
$L_{\rm Bol}\propto T_{\rm X}^2$ 
expected from simple gravitational collapses for formation
of clusters (Kaiser 1986; Navarro, Frenk \& White 1995).

As shown in Section 6 (and Fig. 4), 
the mass estimated using the tentatively-identified lensing arc
and the gravitational mass derived from the X-ray deprojection analaysis 
are in good agreement. 
Moreover, the core radius of $60\pm 20$ kpc measured for 4C+55.16
is similar to the best-fit mean value of $\sim 50$ kpc measured
in the six lensing cooling-flow clusters studied by Allen (1998b).

The inferred metallicity of this cluster gas is $\sim 0.5$\Zs, which 
is not unusual, but is certainly one of the higher values measured
among the ASCA cluster sample compiled by Allen \& Fabian (1998b).
They showed that cooling-flow clusters show higher metallicity
than non cooling-flow clusters, and suggest that the sharply peaked 
X-ray brightness profiles may give
apparently high values of the emissivity-weighted metallicity
in cooling flow clusters 
when there is a metallicity gradient in the cluster core. 
Thus the high metallicity in 4C+55.16 may be due to a steep metallicity
gradient towards the cluster centre.
The high metallicity measured in the spectrum also rules out significant
contribution from the active nucleus to the observed hard X-ray emission
around the Fe K band, otherwise the line would be less prominent.

We conclude that all of the available evidence points to the environment of 
4C+55.16 being like other distant massive cooling flow clusters.
4C+55.16 is yet another powerful radio galaxy surrounded by a strong 
cooling flow. Unlike Cygnus A and 3C295, 4C+55.16 is a compact radio source.
Detailed high spatial resolution observations with AXAF will be required to
determine the interaction between the radio source and the dense
surrounding intracluster medium.
Probably the most comparable cluster to 4C+55.16 is 
PKS\thinspace 0745--19, which also contains a strong cooling flow
and exhibits gravitationally-lensed arcs (Allen, Fabian \& Kneib 1996).
However, the amorphous radio source shows a steep spectrum
and is almost 2 orders of magnitude less powerful thtan 4C+55.16 at 5 GHz.
Massive cooling flows have not been found so far around `pure' 
compact, GPS cores (e.g., O'Dea et al 1996).

The gas pressure at the centre of a cooling flow is
$P= nT\approx 10^7$ \cubcm\ K (Heckman et al 1989).
This leads to a free-free absorption optical depth 
$\tau_{\rm ff}\approx 0.6P^2_7\nu_9^{-2}T_4^{7/2}l_1$ at a frequency 
$10^9\nu_9$ Hz, through a cloud of length $10\thinspace l_1$ pc at pressure
$10^7P_7$ \cubcm\ K and temperature $10^4T_4$ K.
It is therefore plausible that the turnover seen in the radio spectrum
of the compact core at $\sim 1.7$ GHz (Whyborn et al 1985) 
is due to free-free absorption in the 
H$\alpha $ emitting gas close to the nucleus.
The covering fraction in such clouds must however decrease 
rapidly away from the nucleus in order that the radio knot observed
$\sim 100$ pc north-west of the nucleus (Whyborn et al 1985;
Pearson \& Readhead 1988) is not absorbed.
The inverted spectrum of the counter-jet of NGC1275, which also resides
in a cooling flow,
is suspected to be due to free-free absorption on the pc scale
(Vermeulen, Readhead, \& Backer 1994).
Free-free absorption is favoured for the spectral turnover
in GPS sources by Begelman (1997) and Bicknell, Dopita \& O'Dea (1997). 
Part of the observed optical emission-line luminosity from the innermost
part of 4C+55.16 can be expected from shocked gas surrounding
the jets of such a powerful and a relatively young radio source
(e.g., Bicknell et al 1997), as well as from the cooling flow.

\section*{Acknowledgements}

We thank the ASCA and ROSAT teams for their efforts on operation
of the satellites, and the calibration and maintenance of the software.
The optical image we used in Fig. 1 was taken from the Digitized Sky Survey 
which was produced at the Space Telescope Science Institute (ST ScI) under
U.S. Goverment grant NAG W-2166.
We thank Royal Society (ACE, ACF, SE) and PPARC (KI, SWA) for support.

\end{document}